\documentclass[aps,pra,twocolumn,superscriptaddress,tightenlines,longbibliography,doublespace,nofootinbib,amsmath,amssymb]{revtex4-2}

\usepackage[english]{babel}
\usepackage{graphicx,comment}
\usepackage{braket, color, physics}
\usepackage[colorlinks = true, urlcolor = black, citecolor = blue, linkcolor = red]{hyperref}
\newcommand{\half}{\frac{1}{2}}
\usepackage{titlesec}
\usepackage[T1]{fontenc}
\usepackage{lmodern}

\begin{document}
\title{Quantum counterfactuality with identical particles}
\author{Vinod N. Rao}
\email{vinod@ppisr.res.in}
\affiliation{Theoretical Sciences Division, Poornaprajna Institute of Scientific Research, Bidalur, Bengaluru - 562164, India}
\affiliation{Graduate Studies, Manipal Academy of Higher Education, Manipal - 576104, India}
\author{Anindita Banerjee}
\affiliation{Centre of Development of Advanced Computing, Corporate Research and Development, Pune - 411008, India}
\author{R. Srikanth}
\email{srik@ppisr.res.in}
\affiliation{Theoretical Sciences Division, Poornaprajna Institute of Scientific Research, Bidalur, Bengaluru - 562164, India}
\begin{abstract}
Quantum self-interference enables the counterfactual transmission of information, whereby the transmitted bits involve no particles traveling through the channel. In this work, we show how counterfactuality can be realized even when the self-interference is replaced by interference between identical particles. Interestingly, the facet of indistinguishability called forth here is associated with first-order coherence, and is different from the usual notion of indistinguishability associated with the (anti-)commutation relations of mode operators. From an experimental perspective, the simplest implementation of the proposed idea can be realized by slight modifications to existing protocols for differential-phase-shift quantum key distribution or interaction-free measurement.\\

\end{abstract}
\maketitle

\section{Introduction \label{sec:intro}}

\begin{figure}[ht]
\includegraphics[width=\columnwidth]{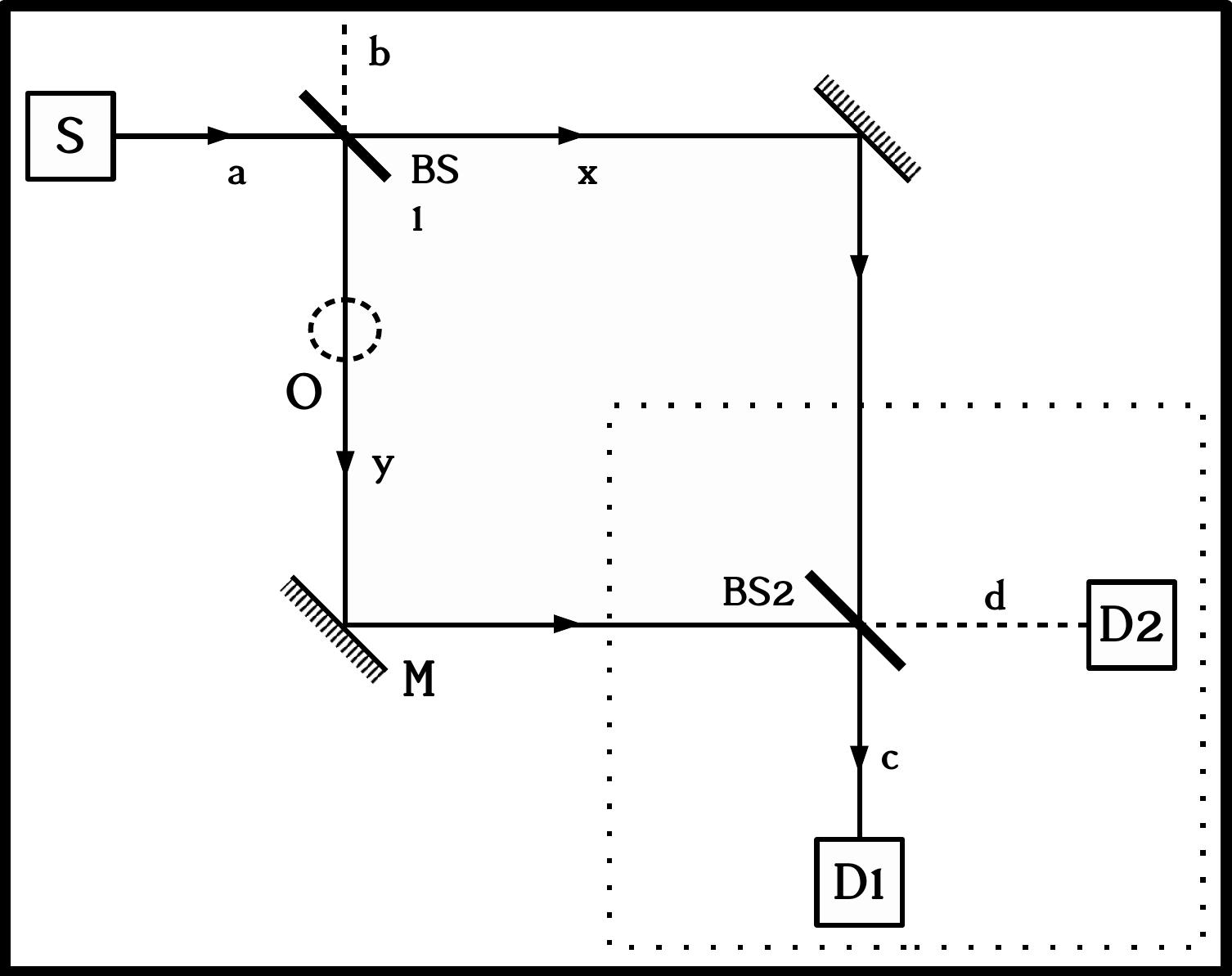}
\caption{Mach-Zehnder interferometer to observe IFM.}
\label{fig:ifm}
\end{figure}

In his classic treatise on quantum mechanics \cite{dirac1981principles}, Dirac famously asserts, ``Each photon then interferes only with itself. Interference between different photons never occurs.'' A simple manifestation of this phenomenon of self-interference is the familiar Mach-Zehnder (MZ) interferometer, where a single photon incident on a beam-splitter BS$_1$ is split into two partial waves, which are re-interfered at the second beam-splitter BS$_2$, leading to a detection solely at detector $D_1$, and a dark fringe at detector $D_2$ (Fig. \ref{fig:ifm}). 
When one of the two interfering paths is blocked, the destructive interference is disrupted, leading to a detection of the photon (with probability $\frac{1}{4}$) at $D_2$. Conditioned on such an event, the presence of the blockade can be inferred deterministically. Interaction-free measurement (IFM) refers to this feature of quantum mechanics that enables detecting the presence of the blockade without interacting with it \cite{elitzur1993quantum}. 

IFM can be applied to quantum key distribution \cite{guo1999quantum, noh2009counterfactual, sun2010counterfactual, shenoy2013semi, rao2021noiseless}, certificate authorization \cite{shenoy2014counterfactual} and entanglement generation \cite{shenoy2015counterfactual}. In these instances, IFM is a basis for counterfactual communication, i.e., communication without particles being transmitted. However, the communication here is not direct, since the \textit{absence} of the obstacle cannot be deterministically inferred by the information recipient, necessitating a public announcement. Starting with Ref. \cite{salih2013protocol}, various authors have studied counterfactual direct communication \cite{salih2013protocol, aharonov2019modification, vaidman2019analysis, hance2021quantum}, and its applications to quantum computation \cite{cao2020counterfactual}, ghost imaging \cite{hance2021counterfactual}, and entangling two geographically separated qubits \cite{guo2015counterfactual}. The nature of counterfactuality of the one of the bit values (corresponding to the obstacle's absence) in these protocols has provoked a lively debate \cite{gisin2013optical, vaidman2014comment, salih2014salih, vaidman2019analysis, aharonov2019modification, hance2021quantum, salih2022laws, salih2021exchange}. For our purpose, it suffices to note that all these protocols involve the self-interference of single photons, consistent with Dirac's requirement on what sets quantum interference apart from the classical one.

\begin{figure}[ht]
\includegraphics[width=\columnwidth]{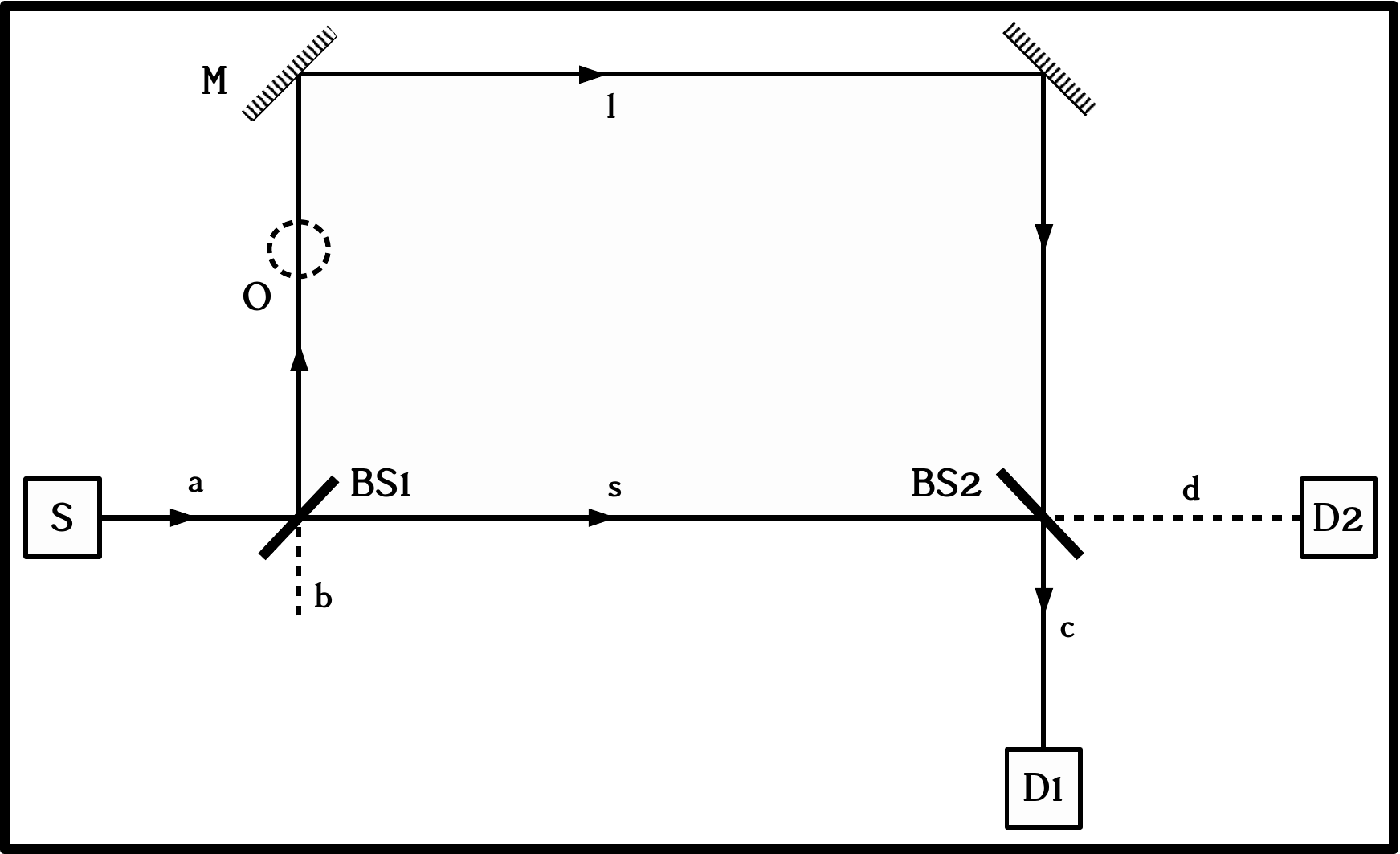}
\caption{Differential phase-shift interferometer: the incident light is in the form of a train of weak coherent pulses. Owing to the difference in interferometric arm length, consecutive particles interfere at beam-splitter BS$_2$. For IFM, we include a retractable obstacle $O$.}
\label{fig:ifm2}
\end{figure}

Yet, Dirac's dictum on self-interference is known today to be too simplistic \cite{glauber1995dirac}, and in particular is implicitly falsified in differential-phase-shift quantum key distribution (DPS QKD) \cite{inoue2003differential,honjo2004differential,takesue2005differential, diamanti2006100} and twin-field (TF) QKD \cite{lucamarini2018overcoming, wang2018twin, cui2019twin}. The DPS interferometer works essentially in the same manner as the MZ interferometer, except that one arm is longer, to ensure that the interfering partial waves at the second beam-splitter belong to two consecutive identical particles (Fig. \ref{fig:ifm2}) of a given optical field. In the TF interferometer, the two interfering particles belong to different optical fields, the basic idea of which is minimally depicted by the dotted box in Fig. \ref{fig:ifm}. Thus the DPS interferometer or TF interferometer implements an interference between two distinct (identical) particles, and yet works just like the self-interference of a single particle to yield MZ-like statistics.

The present work studies the analogue of IFM for the DPS and TF interferometers. This modification conceptually transforms IFM from a phenomenon based on photonic self-interference to one based on interference between two identical particles. The result is an IFM that operates across two particles, i.e., one where a blockade on one particle's path is ascertained without interaction with it, by means of a detection on another, identical particle-- a feature that may appropriately be called ``IFM-by-proxy''. In the following few sections, we will discuss the case of the DPS interferometer, but similar arguments also apply to the TF interferometer and will be specifically discussed in Section \ref{sec:conc}.

\section{DPS interferometer \label{sec:tpifm}}

The DPS interferometer corresponds to a pulse-train adaptation of the MZ interferometer, where one interferometer arm is longer than the other (Fig. \ref{fig:ifm2}). We consider a train of $N$ identical, weak coherent pulses incident on beams-splitter BS$_1$. Consecutive pulses are spaced by a constant interval such that the partial wave of $j$-th pulse traveling via the short arm and that of $(j-1)$-th pulse traveling via the long arm, interfere at beam-splitter BS$_2$. 

The state of the train is given by:
\begin{equation}
\ket{\Psi} = \bigotimes_{j=1}^{N} \ket{\alpha e^{\text{i}\phi_j}},
\label{eq:coherent}
\end{equation}
where $|\alpha|^2$ is the mean photon number of the train and $\phi_j$ is the phase of the $j$th pulse.

Under the action of a beam-splitter BS$_1$, for the $j$-th pulse, we have $\ket{\alpha}_j\ket{\rm vac} \xrightarrow{{\rm BS}_1} \ket{\frac{\alpha}{\sqrt{2}}}_{j,s}\ket{\frac{\text{i}\alpha}{\sqrt{2}}}_{j,l}$. On account of the path difference, the partial waves of two consecutive pulses meet, and we have
\begin{equation}
\ket{\frac{\alpha}{\sqrt{2}}}_{j,s}\ket{\frac{\text{i}\alpha}{\sqrt{2}}}_{j-1,l} \xrightarrow{{\rm BS}_2} \ket{\alpha}_{j,c}\ket{\rm vac}_{j,d},
\label{eq:BS2}
\end{equation}
implying that there can be a detection at detector $D_1$ and none at $D_2$. We note that in Eq. (\ref{eq:BS2}) the interfering partial waves at BS$_2$ belong to two consecutive pulses. Therefore, the quantum indistinguishability of the photons in the pulses is crucial for this interference to happen.

The detection occurs at $D_2$ instead of $D_1$ by letting $\phi_j = \phi_{j-1}+\pi$, i.e., if the phase difference of $\pi$ exists between two consecutive pulses. Thus the pulse-train sender can transmit information by modulating the phase $\phi_j \in \{0,\pi\}$ of each pulse. This fact forms the basis of DPS QKD, where a key bit is generated conditioned on whether detection happens at $D_1$ or $D_2$. For cryptographic security, DPS QKD requires the coherent pulse to be highly attenuated, i.e., $|\alpha|^2\ll 1$, which ensures that the two possible encoding states can be made sufficiently non-orthogonal, as $\bra{\alpha}\ket{-\alpha} = \text{e}^{-2|\alpha|^2}$ \cite{waks2006security, moroder2012security}. 

\section{IFM with a DPS interferometer \label{sec:ifmanalog}}
Our point of departure is the observation that since the DPS interferometer reproduces MZ interferometric statistics (in the regime where DPS QKD is valid), the IFM principle in Figure \ref{fig:ifm} can be applied to the DPS setup of Figure \ref{fig:ifm2}. We show this below.

Accordingly, in Fig. \ref{fig:ifm2}, we may insert the retractable blockade or obstacle $O$ in the long path $l$. If $O$ is inserted, then the pulse amplitude in arm $l$ is blocked. Therefore, in place of Eq. (\ref{eq:BS2}), we have
\begin{equation}
\ket{\frac{\alpha}{\sqrt{2}}}_{j,s}\ket{\rm vac}_{j-1,l} \xrightarrow{{\rm BS}_2} \ket{\frac{\alpha}{2}}_{j,c}\ket{\frac{\text{i}\alpha}{2}}_{j,d},
\label{eq:BS2+}
\end{equation}
showing that there could be a detection at detector $D_2$ with equal probability as at $D_1$. 

We require $|\alpha|^2 \ll 1$ to ensure that with high probability, only a single detection occurs per pulse. Here IFM is realized because we ascertain the presence of $O$ without interacting with it, by means of a detection at $D_2$. Intriguingly, the detection of the $j$-th pulse indicates the blocking of \textit{another} pulse, namely the $(j-1)$-th. It may be apt to refer to this kind of IFM, which invokes quantum indistinguishability, as `IFM-by-proxy'.

Here it is worth mentioning that we obtain IFM-like behavior precisely in the same limit that DPS QKD gives cryptographic security. This is due to the fact that attenuated pulses emulate the nonclassical behavior of single-photons \cite{moroder2012security}. Prima facie, this seems ironic since coherent states show the most classical behavior in one sense, namely that they saturate the preparation uncertainty or information-vs-disturbance trade-off in the quadratures \cite{sabuncu2007nonunity}.

To show why a train of coherent states with sufficient attenuation is necessary to DPS interferometry, consider instead a train of single photons, described by the state:
\begin{equation}
\ket{\Phi} \equiv \bigotimes_{j=1}^{N} \hat{a}^\dagger_j \ket{\rm vac}.
\label{eq:stateX}
\end{equation}
It is at once evident that it does not lead to MZ-like statistics, but instead to a probabilistic Hong-Ou-Mandel effect, which is a two-photon interference phenomenon. 

It is not hard to show that a DPS interferometer ideally requires the ``tensor sum'' train of $N$ single-photons: 
\begin{equation}
\hat{Q}_N\ket{\rm vac} \equiv \frac{1}{\sqrt{N}} \sum_{j=1}^{N} \hat{a}^\dagger_j\ket{\rm vac},
\label{eq:state}
\end{equation}
rather than the ``tensor product'' train of Eq. (\ref{eq:stateX}). The efficacy of the coherent-state train of Eq. (\ref{eq:coherent}) to produce IFM-by-proxy behavior rests on the fact that it approximates Eq. (\ref{eq:state}) in the limit of sufficient attenuation.

In terms of the sum-train of Eq. (\ref{eq:state}), the coherent-state train of Eq. (\ref{eq:coherent}) can be written as
$\ket{\Psi} = \sum_{m=0}^{\infty}\frac{N^{m/2}|\alpha|^m}{\sqrt{m!}} (\hat{Q}_N)^m \ket{\rm vac}.$
Thus, the sum-train state can in principle be engineered from the coherent-state train by means of suitable nonlinear filtering to remove the terms corresponding to $m=0$ (vacuum) and $m>1$ (higher-order excitations).

The sum-train of Eq. (\ref{eq:state}) is of theoretical interest also to show that IFM-by-proxy can be understood using just ``first quantization'' arguments. The action of beam-splitter BS$_1$ is given by $\hat{a}_j \longrightarrow \frac{1}{\sqrt{2}}(\hat{l}_j + \text{i}\hat{s}_j)$, and that of BS$_2$ similarly acting on input modes $l_j$ and $s_j$. Then, the electric field operators corresponding to detectors $D_1$ and $D_2$ are given by (up to a global phase)
$
\hat{c}^{\dagger}_j(t_j) =\frac{1}{\sqrt2}(\hat{l}_{j-1}\text{e}^{\text{i}(k\delta - \omega t_{j-1})} + \text{i}\hat{s}_j \text{e}^{-\text{i} \omega t_{j}})$ and
$\hat{d}^{\dagger}_j(t_j) = \frac{1}{\sqrt2}(\text{i}\hat{l}_{j-1}\text{e}^{\text{i}(k\delta - \omega t_{j-1})} + \hat{s}_j \text{e}^{-\text{i} \omega t_{j}})$,
where $\omega, \delta$ and $k$ denote the angular frequency, inter-particle distance and wave number, respectively. It follows that the probability of detection of a photon at detector $D_1$ at time $t_j$ is
\begin{equation}
\langle \hat{c}_j \hat{c}^{\dagger}_j \rangle = \half (1 + \cos\left[k\delta - \omega (t_{j-1} - t_{j})\right]).
\label{eq:d1expt}
\end{equation} 
Since $k\delta = \omega (t_{j-1} - t_{j})$, this interference is like that in a conventional MZ interferometer, except that the two partial waves that converge at BS$_2$ belong to two distinct (consecutive), indistinguishable pulses. Although Eq. (\ref{eq:d1expt}) does not correspond to self-interference but instead to interference between two photons, yet it involves first-order (rather than higher-order) coherence, and in that sense is a kind of \textit{single-photon} interference. 

Furthermore, the type of indistinguishability we invoke to obtain Eq. (\ref{eq:d1expt}) may be differentiated from the ``conventional'' one between identical photons that is associated with genuine two-photon or multi-photon interference, and is responsible for such effects as the Hong-Ou-Mandel effect \cite{hong1987measurement}, bosonic stimulation \cite{shenoy2013efficient}, and the quantum advantage in boson sampling \cite{aaronson2011computational}. In the conventional case, indistinguishability is imposed by the commutation relations such as $[\hat{l}_j^\dagger,\hat{s}_j^\dagger]=0$. However, to explain IFM-by-proxy with the state Eq. (\ref{eq:state}), this relation is not required, essentially because this state involves no quadratic or higher-order operators. The indistinguishability here can be understood as a manifestation of first-order coherence between the incoming photonic fields.

\section{Squeezed-light pulse train}
We know that the attenuated coherent state train is conducive to exhibit DPS interferometry, whereas the product-train state is not. This brings up the question of general characteristics of a pulse train to yield DPS interferometry, and by extension IFM-by-proxy. We address this now.

It is particularly instructive to study a train of squeezed light, which has been the subject of extensive experimental studies \cite{lassen2010experimental}. Consider a train of pulses in the squeezed (vacuum) state, given by
$\ket{\zeta} = \hat{S}(\eta)\ket{\rm vac} \equiv {\rm exp} ({-(\eta/2)\hat{a}^{\dagger2} + (\eta^\ast/2) \hat{a}^2})\ket{\rm vac} = \sqrt{ \sech ~z} ~\sum_{n=0}^{\infty} \frac{\sqrt{(2n)!}}{n!} \bigg(\frac{\text{e}^{-\text{i}\phi}}{2} ~(\tanh ~z) \bigg)^n \ket{2n},$ 
where $\hat{S}(\eta)$ is the squeeze operator and $\eta = z\text{e}^{\text{i} \phi}$ is the squeezing parameter. For $z \ll 1$, the attenuated squeezed-light train yields a two-photon train in place of Eq. (\ref{eq:state}), and thus is unsuitable for DPS interferometry.

This relates to the fact that the squeezed-light train does not reproduce MZ interferometric behavior. Given the $j$th pulse incident on a beam-splitter, the output is the mode-entangled state
$B_{(1)} \hat{S}_{j}(\eta)\ket{\rm vac} = \hat{S}_{j;s}(\eta/2) \hat{S}_{j;l}(\eta/2) 
\hat{S}_{j;sl}(\eta/2)\ket{\rm vac},$
where the two-mode squeeze operator $\hat{S}_{j;sl}$ indicates the entanglement between the modes $s$ and $l$ corresponding to the $j$th pulse. 

In the case of Fig. \ref{fig:ifm2}, at BS$_2$, the mode $(j;l)$ does not meet its entangled counterpart mode $(j;s)$, but instead mode $(j+1,s)$, which for its part is entangled with mode $(j+1,l)$. Owing to the absence of entanglement between the two incoming modes, the average photon number detected at the output mode $\hat{d}_{j}$ of BS$_2$ at time $t_j$, is non-zero. To show this, letting
$\ket{\Psi(S)} \equiv \cdots \hat{S}_{j+1}(\eta)\hat{S}_j(\eta) \cdots \ket{\rm vac},$
we find that the average photon number $\langle \hat{d}^\dagger_j \hat{d}_j \rangle$ detected at time $t_j$ at detector $D_2$ is
\begin{subequations}
\begin{align}
\langle \hat{d}_{j}^\dagger \hat{d}_{j} \rangle &= 
\half \langle (\text{i}\hat{l}_{j}^\dagger + s_{j+1}^{\dagger})(-\text{i}\hat{l}_{j} + \hat{s}_{j+1}) \rangle \\
&= \frac{1}{4}\langle (\text{i}\hat{a}_{j}^\dagger + \hat{a}_{j+1}^{\dagger})(-\text{i}\hat{a}_{j} + \hat{a}_{j+1}) \rangle \label{eq:l2a} \\
&= \frac{1}{4}\left[ 2\sinh^2(z) +\text{i} \langle \hat{a}_{j}^\dagger \hat{a}_{j+1}\rangle -\text{i} \langle \hat{a}_{j+1}^\dagger \hat{a}_{j}\rangle\right]
\label{eq:aa} \\
&= \half \sinh^2(z) \label{eq:bb}.
\end{align}
\label{eq:stuff}
\end{subequations}
Operators $\hat{b}_{j}$ and $\hat{b}_{j+1}$ to the input modes of BS$_1$ are conveniently dropped in Eq. (\ref{eq:l2a}) as $\hat{b}_{j}\ket{\Psi} = \hat{b}_{j+1}\ket{\Psi} = 0$. To obtain Eq. (\ref{eq:aa}), we have used the identities $\hat{S}^\dagger(\eta) \hat{a} \hat{S}(\eta) = \hat{a} \cosh(z) - \hat{a}^\dagger \text{e}^{\text{i} \phi} \sinh(z)$, $\hat{S}^\dagger(\eta) a^\dagger \hat{S}(\eta) = \hat{a}^\dagger \cosh(z) - \hat{a} \text{e}^{-\text{i} \phi} \sinh(z)$, and finally the relations $[\hat{a}_{j},\hat{a}_{j+1}] = [\hat{a}_{j},\hat{a}_{j+1}^\dagger] = 0$ to obtain Eq. (\ref{eq:bb}).

Eq. (\ref{eq:stuff}) entails that $\langle \hat{d}^\dagger_j \hat{d}_j \rangle > 0$ for $z>0$. In other words, we do not obtain MZ-like statistics, and thus IFM-by-proxy is ruled out. Evidently, this is due to the two-mode entanglement produced by the beam splitter's action. 
If instead of the squeezed train $\ket{\Psi(S)}$, the train of coherent pulses of Eq. (\ref{eq:coherent}) is considered, then in place of Eq. (\ref{eq:stuff}) and noting the identity $\hat{a}\ket{\alpha} = \alpha\ket{\alpha}$, we have $\langle \hat{d}_{j}^\dagger \hat{d}_{j} \rangle = \frac{1}{4}\langle (a_{j}^\dagger - a_{j+1}^{\dagger})(a_{j} - a_{j+1}) \rangle = \half(|\alpha|^2-|\alpha|^2)=0$, as expected.

\section{IFM with multiple particles \label{sec:3pulse}}

\begin{figure}[ht]
\includegraphics[width=\columnwidth]{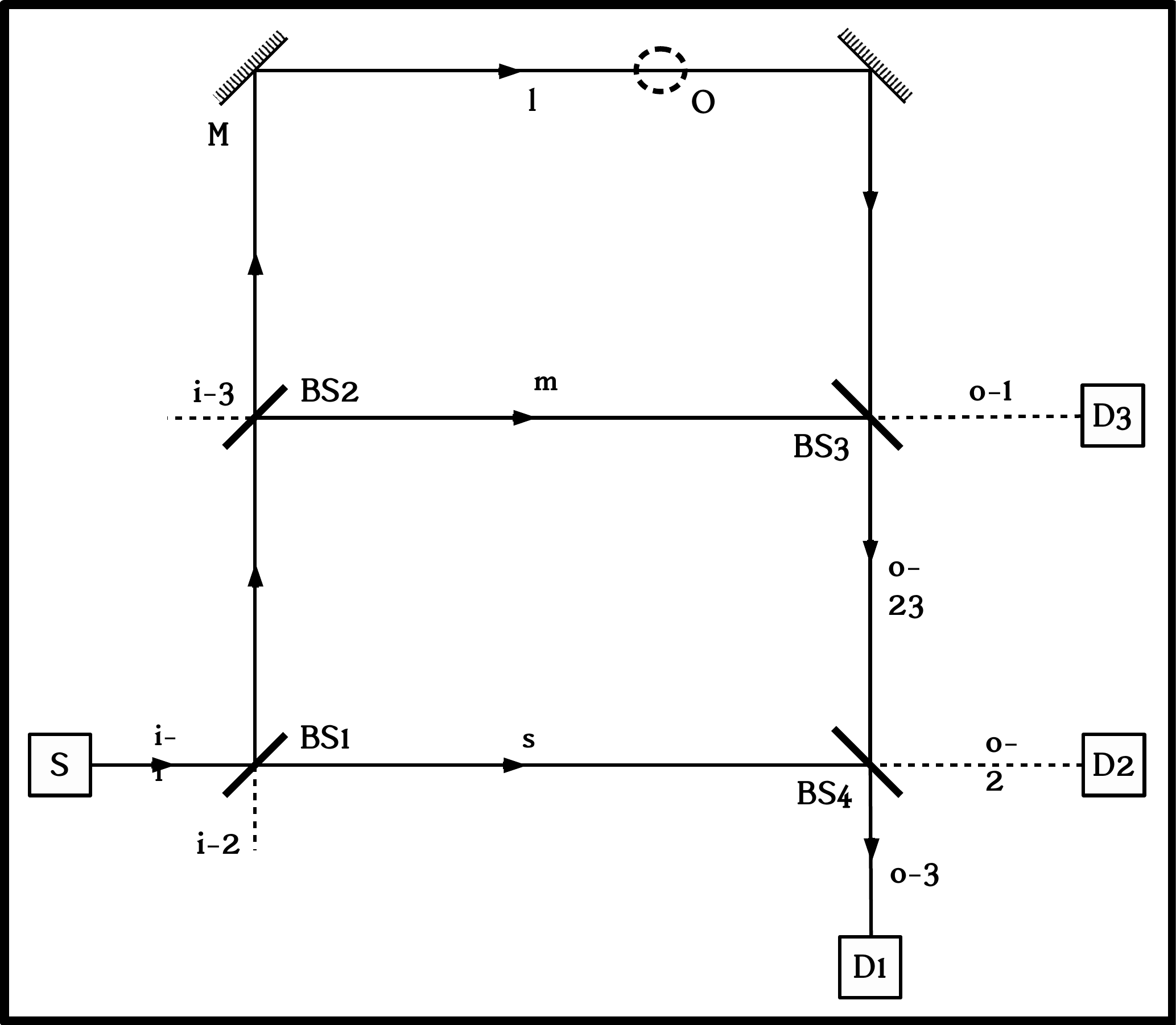}
\caption{Schematic of an interferometric setup to observe IFM-by-proxy with three consecutive pulses. Without the obstacle $O$, the detection event happens only at detector $D_1$ alone. In the presence of the obstacle in either arm $l$ or $m$, a detection at detectors $D_2$ or $D_3$ is possible, which constitutes IFM-by-proxy. Here the path labels $k{-}n$ indicate the respective input or output ports.}
\label{fig:ifm3}
\end{figure}

The principle of IFM-by-proxy can be straightforwardly extended to a situation where multiple pulses interfere in an interferometric setup. Consider the case of an analogous single-photon interference in a 3-pulse scenario. For example, the beam-splitters of Fig. \ref{fig:ifm2} are replaced by tritters (three-way beam-splitters) described by the unitary 
\begin{equation}
\mathbf{U}_3 = \begin{pmatrix}
\frac{1}{\sqrt{2}} & \frac{\text{i}}{2} & -\frac{1}{2} \\
\frac{\text{i}}{\sqrt{2}} & \frac{1}{2} & \frac{\text{i}}{2} \\
 0 & \frac{\text{i}}{\sqrt{2}} & \frac{1}{\sqrt{2}}
 \end{pmatrix}.
 \label{eq:3way}
 \end{equation}
In general, any $n$-input $n$-output splitter can be realized by a cascaded setup of $\mathbf{U}_2$ \cite{reck1994experimental}. The tritter transformation of Eq. (\ref{eq:3way}) realized through a two-interferometer cascaded setup is depicted in Fig. \ref{fig:ifm3}.

In the absence of a retractable obstacle $O$, only a $D_1$ detection occurs for the train of pulses given in Eq. (\ref{eq:coherent}). To show this, note that the state of the fields after BS$_2$ is given by:
\begin{align}
\ket{\psi} &= \ket{\frac{\alpha}{\sqrt{2}}}_{j,s} \ket{\frac{-\alpha}{2}}_{j-1,m} \ket{\frac{\text{i}\alpha}{2}}_{j-2,l} \nonumber \\
&\xrightarrow{{\rm BS}_3} \ket{\frac{\alpha}{\sqrt{2}}}_{j,s} \ket{\frac{-\alpha}{\sqrt{2}}}_{\ast,{\rm o-23}} \ket{\rm vac}_{\ast,{\rm o-1}} \nonumber \\
&\xrightarrow{{\rm BS}_4} \ket{-\alpha}_{\ast,{\rm o-3}} \ket{\rm vac}_{\ast,{\rm o-2}} \ket{\rm vac}_{\ast,{\rm o-1}}, 
\label{eq:3train}
\end{align}
where $\ast$ indicates the superposition of two or more consecutive pulses. From Eq. (\ref{eq:3train}), it follows that a detection at detector $D_2$ or $D_3$ indicates the presence of $O$ on path $l$-- a case of IFM.

A potential application of this kind of setup is checking for misalignment, defects or environment-induced decoherence in a quantum circuit. One can run an error diagnostics on the circuit by introducing obstacles or certain phase fluctuations and monitoring the detector clicks. Quantum information leaking into the interferometer's environment can be compensated via error correcting codes or recovered using the leaked modes and a classical feed-forward technique \cite{sabuncu2010environment}.

Another type of chaining, using a large number of interferometers, can lead to asymptotically transmit an exclusive bit, i.e., information not only of the presence of the obstacle, but also of its absence. The counterfactuality here arises from the chained Zeno effect, as discussed for slightly different setups in Refs. \cite{salih2013protocol, vaidman2019analysis, hance2021quantum}. To show that these results can be reproduced in a DPS context, we require two observations concerning the transmission of state $\ket{\alpha}$ through a cascade of $n$ beam-splitters. 

(a) First is that the above chained Zeno experiments can be reproduced using coherent states instead of single photons, conditioned on a single-photon detection. Let the transformation across BS be $\ket{1}_{i1}\ket{0}_{i2} = (\cos \theta \ket{10} + \sin \theta \ket{01})_{o1,o2}$ and $\ket{0}_{i1}\ket{1}_{i2} = (\cos \theta \ket{01} - \sin \theta \ket{10})_{o1,o2}$ (where $iP/oP$ indicates the input/output port $P$).
\begin{figure}[ht]
\includegraphics[width=\columnwidth]{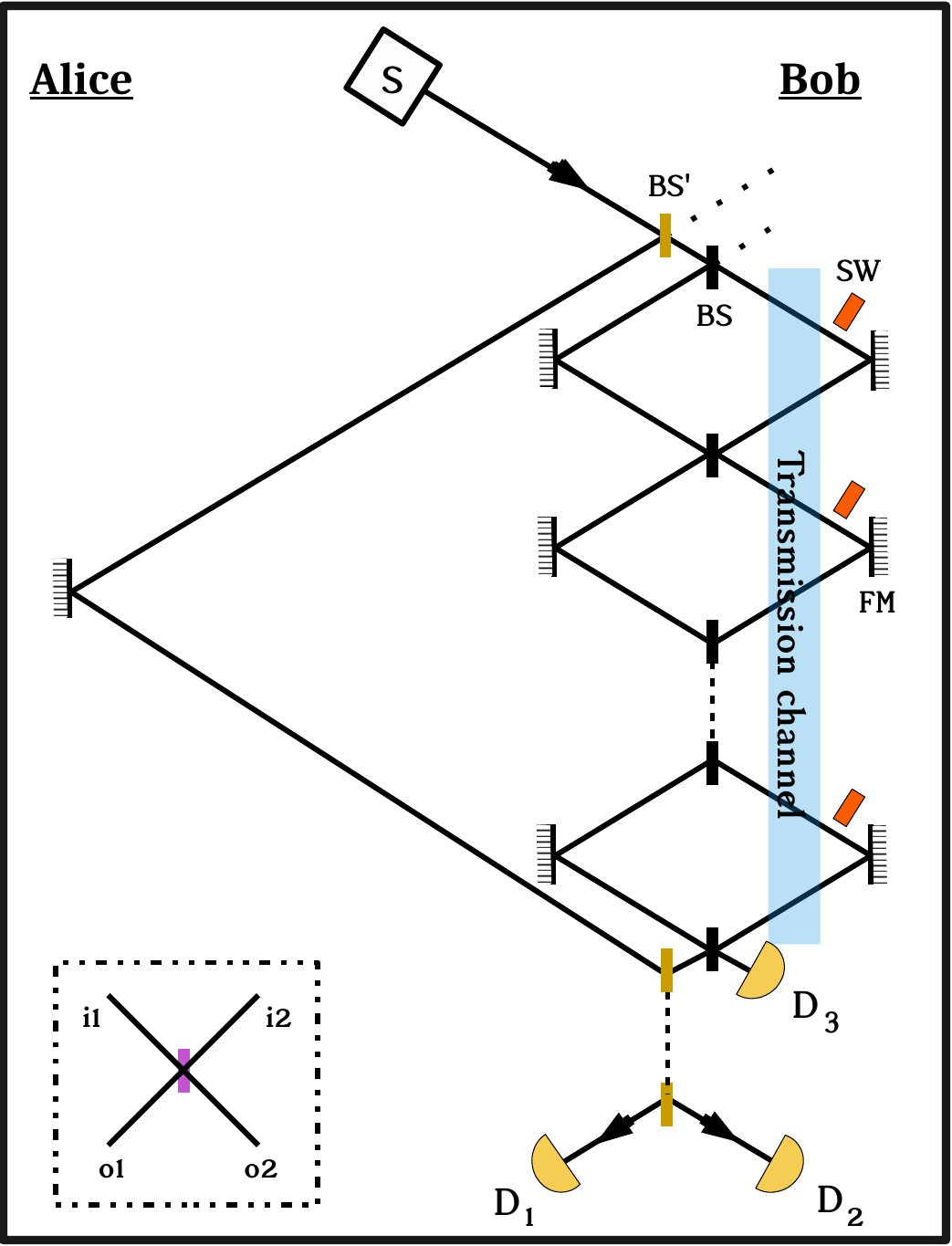}
\caption{Schematic of counterfactual communication using the chained quantum Zeno setup: The outer chain consists of $m$ cascaded beam-splitters BS$^\prime$, and each inner chain consists of $n$ cascaded beam-splitters BS. The inset gives the pattern of input and output ports at each beam-splitter. S - source; BS - beam-splitter; SW - switch; FM - Faraday mirror.}
\label{fig:chained}
\end{figure}

For the case where Bob applies the reflect operation at all interferometers, we find
\begin{align}
\ket{\psi} &= \ket{\alpha}\ket{0} \xrightarrow{\text{BS}_1} \ket{(\cos \theta) \alpha} \ket{(\sin \theta) \alpha} \xrightarrow{\text{BS}_2} \hdots \nonumber \\
&\hdots \xrightarrow{\text{BS}_\text{n}} \ket{(\cos n\theta) \alpha}_{(D2)} \ket{(\sin n\theta) \alpha}_{(D1)},
\label{eq:reflect}
\end{align}
indicating a detection at detector $D_1$ or $D_2$, setting $\theta = \frac{\pi}{2n}$. In the case when Bob applies the blocking operation at all interferometers, we find
\begin{align}
\ket{\psi} &= \ket{\alpha}\ket{0} \xrightarrow{\text{BS}_1} \ket{(\cos \theta) \alpha} \ket{\rm vac} \xrightarrow{\text{BS}_2} \hdots \nonumber \\
& \hdots \xrightarrow{\text{BS}_\text{n}} \ket{(\cos^n \theta) \alpha}_{(D2)} \ket{(\cos^{n-1} \theta \sin \theta) \alpha}_{(D1)},
\label{eq:block}
\end{align} 
indicating a detection at detector $D_3$ with probability approaching 1 for very large $n$. Eqs. (\ref{eq:reflect}) and (\ref{eq:block}) imply that the detection probability at detector $D_1$/$D_2$ and $D_3$, respectively, is asymptotically close to unity, conditioned on the detection of a photon (with probability $|\alpha|^2$). Counterfactuality is evident in the case of Eq. (\ref{eq:block}). In Eq. (\ref{eq:reflect}), only the ``inner chain'' of the cascade is given. With inclusion of the ``outer chain'', as depicted in Fig. \ref{fig:chained}, one can conditionally have fully counterfactual communication according to the respective criterion, along the lines of the schemes presented in Refs. \cite{salih2013protocol, vaidman2019analysis, hance2021quantum}. Here it may be stressed that the coherent states are taken in the single-photon detection and interference limit, so that the criterion for nonclassicality of the counterfactual effect \cite{hance2021quantum} is indeed met.

(b) The second observation is that, if a train of pulses is injected, then owing to quantum indistinguishability the probabilities in Eqs. (\ref{eq:reflect}) and (\ref{eq:block}) are unaffected if optical delay lines are introduced after one or more beam splitters of the inner cycle. If a 1-step delay line is introduced just after BS$_1$, then downstream of BS$_2$, the evolution of the system will not be restricted to the self-interference of each pulse, but the interference between two consecutive. Introducing a second delay line after BS$_2$, we interfere three consecutive pulses at BS$_3$, and so forth. The result is a chaining of DPS interferometers, and generalizes the two-particle interference of the original DPS scheme.

\section{Discussion and Conclusion \label{sec:conc}}

We now briefly summarize the novelty of the work, its potential experimental realization and further ramifications.

\paragraph{Counterfactuality without self-interference.}
At the heart of interaction-free measurement (IFM), as originally introduced, is quantum self-interference and, specifically, its modification under the blocking of an interfering path. The present work proposes a fundamental alteration to this theme, by implementing IFM with the interference between two distinct but identical particles. Thereby we ascertain the presence of an obstacle in the path of a particle in an interaction-free manner, by means of a detection on another particle, located elsewhere. To distinguish this kind of IFM from the conventional IFM based on self-interference, we describe it as an IFM-by-proxy.

In IFM-by-proxy, although two (or more) photons interfere, the result is a single-photon interference effect and can be fully described as a first-order coherence employing only the first-quantization formalism. It does not constitute a two-photon interference, such as occurs in the Hong-Ou-Mandel effect. The aspect of indistinguishability called into play here does not correspond to the exchange symmetry between two or more identical particles, but rather a kind of photonic non-individuality of the particles belonging to a field mode. 

\paragraph{Towards an experimental realization.}
The experimental setup of Fig. \ref{fig:ifm2} to realize IFM-by-proxy is a straightforward extension of that for DPS QKD \cite{honjo2004differential, takesue2005differential, diamanti2006100}, as well as that for IFM \cite{kwiat1995interaction}, and thus well within the scope of current technology. For completeness, we provide a brief description. For the IFM setup, the single photon source is replaced by a source of attenuated coherent states and an optical delay is introduced in one of the arms. For the DPS setup, a removable obstacle is placed in one of the arms. The source could be a continuous wave (cw) laser diode equipped with an external cavity of 810 nm. This can be converted into a pulsed light source by means of a high-speed amplitude modulator placed just behind the cw light source \cite{zhang2009megabits}. We may as well employ 1550 nm telecom wavelength, which may be considered mainly for the availability of InGaAs detectors. However, SiAPD may be preferable thanks to its better performance, specifically its higher efficiency of $70\% $, lower dead-time of 50 ns, etc \cite{takesue200610}. The degree of high attenuation would ensure that there is a much higher probability for single-photon events over multi-photon events. 

Here, it is important to note the effect on the stability due to the unequal path lengths of the interferometer arms in Fig. \ref{fig:ifm2}, and the consequent potentially different degrees of dispersion (such as group-velocity or polarization dispersion) in the two arms. The visibility of the two-photon interference that leads to a probabilistic Hong-Ou-Mandel effect in the context of Eq. (\ref{eq:stateX}), can be used to probe the stability of the modified MZ interferometer \cite{zhao2022propagation, ou2022unbalanced}.

\paragraph{IFM with a twin-field interferometer.} It is not hard to see that IFM-by-proxy will work even in a twin-field (TF) setup, i.e., one where the two interfering particles belong to two distinct optical fields (dotted box in Fig. \ref{fig:ifm}), provided the fields are combined indistinguishably and attenuated sufficiently to ensure single-photon interference. In TF-QKD, pairs of attenuated, phase-randomized optical fields are prepared by two distant parties (Alice and Bob) and sent to Charlie at a central station, where they are interfered and measured (dotted box of Fig. \ref{fig:ifm}). Key bits can be distilled from pairs of fields imparted with an identical random phase. Charlie can know whether Alice and Bob share equal or different bits, but not their absolute bit values. In that sense, TF-QKD provides a natural realization of the measurement-device independent (MDI) \cite{lo2012measurement} version of DPS QKD protocol (or, equivalently, TF-QKD may be considered as the two-sender extension of DPS QKD.). This may be distinguished from an earlier MDI version \cite{ferenczi2013security} of the DPS protocol, which requires two-photon interference to generate the key bits. 

The IFM-by-proxy for the TF interferometer is implicitly realized in the side-channel-free (SCF) QKD protocol \cite{wang2019practical}, which is a modified send-or-not-send (SNS) protocol \cite{wang2018twin}, a variant of the TF-QKD protocols \cite{lucamarini2018overcoming, cui2019twin}. In SCF QKD, Alice and Bob randomly may send or not send an optical field, which is equivalent to a not-blocking or blocking action on a transmitted pulse. Thus, if precisely one of them sends a pulse, then conditioned on Charlie announcing a detection at a single detector, Alice and Bob may infer a secret bit in an MDI manner. Unlike in the original SNS protocol, here Alice and Bob do not randomize the pulse phase. Thus the protocol is sensitive to alignment errors like the TF protocol. However, the use of a single basis of encoding is exploited to eliminate side channels at the source, and thereby minimize leakage of preparation information to an eavesdropper. Evidently, SCF QKD completes the TF analogues of IFM or DPS protocols, as depicted in Table \ref{tab:IFM}.

\begin{table}[h]
\centering
\begin{tabular}{|c|c|c|}
\hline
Interferometry & Interference between & Effect with obstacle \\
\hline
Mach-Zehnder & photon and itself & IFM \\
 & (self-interference) & \\
\hline
DPS QKD & two pulses of the & IFM-by-proxy \\
 & same optical field & \\
\hline
TF-QKD & two pulses of & IFM-by-proxy \\
 & different optical fields & (SCF QKD) \\
\hline
\end{tabular}
\caption{Underlying unity of the interference principle behind IFM, IFM-by-proxy in a DPS setup, and modified SNS TF-QKD.}
\label{tab:IFM}
\end{table}

The key commonality among the TF-QKD protocols (including modified SNS) and DPS QKD is that they employ optical fields, with single-photon interference and single-photon measurement, rather than single-photons per se.

\paragraph{Limitations and future prospects.}
Certain limitations in the proposal may be pointed out. One is the difficulty in engineering the filtering operation to realize the ideal `IFM-by-proxy'. Another is that its diagnostic capacity (as in the context of Fig. \ref{fig:ifm3}) may suffer from multiple practical factors that undermine the counterfactual effect.

A foundational question opened up by our work would be that of the fermionic counterpart of the IFM-by-proxy. Furthermore, it is known that multiphoton, linear interference can be the basis of a powerful, albeit non-universal, model of quantum computing that can efficiently solve the boson sampling problem, which is known to be \#P-hard \cite{aaronson2011computational}. By contrast, the single-photon interference that leads to IFM-by-proxy is not expected to give a greater-than-quadratic speedup, as in Grover search \cite{bennett1997strengths}, since it can be described by first-quantization principles. 

While the present work introduces a combination of counterfactuality with indistinguishability, one may consider other nonclassical features of quantum mechanics that may be so combined. Here we may mention a recent work that studies the effect of quantum Cheshire cat dynamics on counterfactual communication \cite{aharonov2021dynamical}. 

\acknowledgements

V.N.R. and R.S. acknowledge the support from Interdisciplinary Cyber Physical Systems (ICPS) program of the Department
of Science and Technology (DST), India, Grant No. DST/ICPS/QuST/Theme-1/2019/14. V.N.R. thanks U. Shrikant for discussions and, acknowledges the support and encouragement from Admar Mutt Education Foundation.

\bibliography{references}

\end{document}